# Towards Effective Bug Triage with Software Data Reduction Techniques

Jifeng Xuan, He Jiang, *Member*, *IEEE*, Yan Hu, Zhilei Ren,
Weiqin Zou, Zhongxuan Luo, Xindong Wu, *Fellow*, *IEEE*

**Abstract**—Software companies spend over 45 percent of cost in dealing with software bugs. An inevitable step of fixing bugs is bug triage, which aims to correctly assign a developer to a new bug. To decrease the time cost in manual work, text classification techniques are applied to conduct automatic bug triage. In this paper, we address the problem of data reduction for bug triage, i.e., how to reduce the scale and improve the quality of bug data. We combine instance selection with feature selection to simultaneously reduce data scale on the bug dimension and the word dimension. To determine the order of applying instance selection and feature selection, we extract attributes from historical bug data sets and build a predictive model for a new bug data set. We empirically investigate the performance of data reduction on totally 600,000 bug reports of two large open source projects, namely Eclipse and Mozilla. The results show that our data reduction can effectively reduce the data scale and improve the accuracy of bug triage. Our work provides an approach to leveraging techniques on data processing to form reduced and high-quality bug data in software development and maintenance.

——————————— ◆ ———————————

## 1 INTRODUCTION

MINING software repositories is an interdisciplinary domain, which aims to employ data mining to deal with software engineering problems [22]. In modern software development, software repositories are large-scale databases for storing the output of software development, e.g., source code, bugs, emails, and specifications. Traditional software analysis is not completely suitable for the large-scale and complex data in software repositories [58]. Data mining has emerged as a promising means to handle software data (e.g., [7], [32]). By leveraging data mining techniques, mining software repositories can uncover interesting information in software repositories and solve real-world software problems.

A *bug repository* (a typical software repository, for storing details of bugs), plays an important role in managing software bugs. Software bugs are inevitable and fixing bugs is expensive in software development. Software companies spend over 45 percent of cost in fixing bugs [39]. Large software projects deploy bug repositories (also called *bug tracking systems*) to support information collection and to assist developers to handle bugs [14], [9]. In a bug repository, a bug is maintained as a *bug report*, which records the textual description of reproducing the bug and updates according to the status of bug fixing [64]. A bug repository provides a data platform to support many types of tasks on bugs, e.g., fault prediction [7], [49], bug localization [2], and reopened-bug analysis [63]. In this paper, bug reports in a bug repository are called *bug data*.

There are two challenges related to bug data that may affect the effective use of bug repositories in software development tasks, namely the large scale and the low quality. On one hand, due to the daily-reported bugs, a large number of new bugs are stored in bug repositories. Taking an open source project, Eclipse [13], as an example, an average of 30 new bugs are reported to bug repositories per day in 2007 [3]; from 2001 to 2010, 333,371 bugs have been reported to Eclipse by over 34,917 developers and users [57]. It is a challenge to manually examine such large-scale bug data in software development. On the other hand, software techniques suffer from the low quality of bug data. Two typical characteristics of low-quality bugs are noise and redundancy. Noisy bugs may mislead related developers [64] while redundant bugs waste the limited time of bug handling [54].

A time-consuming step of handling software bugs is bug triage, which aims to assign a correct developer to fix a new bug [1], [25], [3], [40]. In traditional software development, new bugs are manually triaged by an expert developer, i.e., a human triager. Due to the large number of daily bugs and the lack of expertise of all the bugs, manual bug triage is expensive in time cost and low in accuracy. In manual bug triage in Eclipse, 44 percent of bugs are assigned by mistake while the time cost between opening one bug and its first triaging is 19.3 days on average [25]. To avoid the expensive cost of manual bug triage, existing work [1] has proposed an automatic bug triage approach, which applies text classification techniques to predict developers for bug reports. In this approach, a bug report is mapped to a document and a related developer is mapped

————————————————

- *J. Xuan is with the School of Software, Dalian University of Technology, Dalian, China, and INRIA Lille – Nord Europe, Lille, France. E-mail: xuan@mail.dlut.edu.cn.*
- *H. Jiang, Y. Hu, Z. Ren, and Z. Luo are with the School of Software, Dalian University of Technology, Dalian, China. E-mail: hejiang@ieee.org, {huyan, zren, zxluo}@dlut.edu.cn.*
- *W. Zou is with Jiangxi University of Science and Technology, Nanchang, China. E-mail: weiqinzou315@gmail.com.*
- *X. Wu is with the School of Computer Science and Information Engineering, Hefei University of Technology, Hefei, China, and Department of Computer Science, University of Vermont, USA. E-mail: xwu@uvm.edu.*

*Manuscript received (insert date of submission if desired).*

to the label of the document. Then, bug triage is converted into a problem of text classification and is automatically solved with mature text classification techniques, e.g., Naive Bayes [12]. Based on the results of text classification, a human triager assigns new bugs by incorporating his/her expertise. To improve the accuracy of text classification techniques for bug triage, some further techniques are investigated, e.g., a tossing graph approach [25] and a collaborative filtering approach [40]. However, large-scale and low-quality bug data in bug repositories block the techniques of automatic bug triage. Since software bug data are a kind of free-form text data (generated by developers), it is necessary to generate well-processed bug data to facilitate the application [66].

In this paper, we address the problem of data reduction for bug triage, i.e., how to reduce the bug data to save the labor cost of developers and improve the quality to facilitate the process of bug triage. Data reduction for bug triage aims to build a small-scale and high-quality set of bug data by removing bug reports and words, which are redundant or non-informative. In our work, we combine existing techniques of instance selection and feature selection to simultaneously reduce the bug dimension and the word dimension. The reduced bug data contain fewer bug reports and fewer words than the original bug data and provide similar information over the original bug data. We evaluate the reduced bug data according to two criteria: the scale of a data set and the accuracy of bug triage. To avoid the bias of a single algorithm, we empirically examine the results of four instance selection algorithms and four feature selection algorithms.

Given an instance selection algorithm and a feature selection algorithm, the order of applying these two algorithms may affect the results of bug triage. In this paper, we propose a predictive model to determine the order of applying instance selection and feature selection. We refer to such determination as prediction for reduction orders. Drawn on the experiences in software metrics[1], we extract the attributes from historical bug data sets. Then, we train a binary classifier on bug data sets with extracted attributes and predict the order of applying instance selection and feature selection for a new bug data set.

In the experiments, we evaluate the data reduction for bug triage on bug reports of two large open source projects, namely Eclipse and Mozilla. Experimental results show that applying the instance selection technique to the data set can reduce bug reports but the accuracy of bug triage may be decreased; applying the feature selection technique can reduce words in the bug data and the accuracy can be increased. Meanwhile, combining both techniques can increase the accuracy, as well as reduce bug reports and words. For example, when 50% bug reports and 70% words are removed, the accuracy of Naive Bayes on Eclipse improves by 2% to 12% and the accuracy on Mozilla improves by 1% to 6%. Based on the attributes from historical bug data sets, our predictive model can provide the accuracy of 71.8% for predicting the reduction order. Based on top node analysis of the attributes, results show that no individual attribute can determine the reduction order and each attribute is helpful to the prediction.

The primary contributions of this paper are as follows.

1. We present the problem of data reduction for bug triage. This problem aims to augment the data set of bug triage in two aspects, namely 1) to simultaneously reduce the scales of the bug dimension and the word dimension and 2) to improve the accuracy of bug triage.

2. We propose a combination approach to addressing the problem of data reduction. This can be viewed as an application of instance selection and feature selection in bug repositories.

3. We build a binary classifier to predict the order of applying instance selection and feature selection. To our knowledge, the order of applying instance selection and feature selection has not been investigated in related domains.

This paper is an extension of our previous work [62]. In this extension, we add new attributes extracted from bug data sets, prediction for reduction orders, and experiments on four instance selection algorithms, four feature selection algorithms, and their combinations.

The remainder of this paper is organized as follows. Section 2 provides the background and motivation. Section 3 presents the combination approach for reducing bug data. Section 4 details the model of predicting the order of applying instance selection and feature selection. Section 5 presents the experiments and results on bug data. Section 6 discusses limitations and potential issues. Section 7 lists the related work. Section 8 concludes.

## 2 BACKGROUND AND MOTIVATION

### 2.1 Background

Bug repositories are widely used for maintaining software bugs, e.g., a popular and open source bug repository, Bugzilla [5]. Once a software bug is found, a *reporter* (typically a developer, a tester, or an end user) records this bug to the bug repository. A recorded bug is called a *bug report*, which has multiple items for detailing the information of reproducing the bug. In Fig. 1, we show a part of bug report for bug 284541 in Eclipse[2]. In a bug report, the summary and the description are two key items about the information of the bug, which are recorded in natural languages. As their names suggest, the summary denotes a general statement for identifying a bug while the description gives the details for reproducing the bug. Some other items are recorded in a bug report for facilitating the identification of the bug, such as the product, the platform, and the importance. Once a bug report is formed, a human triager assigns this bug to a developer, who will try to fix this bug. This developer is recorded in an item assigned-to. The assigned-to will change to another developer if the previously assigned developer cannot fix this bug. The process of assigning a correct developer for fix-ing the bug is called *bug triage*. For example, in Fig. 1, the developer Dimitar Giormov is the final assigned-to developer of bug 284541.

---

[1] The subject of software metrics denotes a quantitative measure of the degree to software based on given attributes [16]. Existing work in software metrics extracts attributes from an individual instance in software repositories (e.g., attributes from a bug report) while in our work, we extract attributes from a set of integrated instances (e.g., attributes from a set of bug reports). See Section S1 in the supplemental material, http://oscar-lab.org/people/~jxuan/reduction/.

[2] Bug 284541, https://bugs.eclipse.org/bugs/show_bug.cgi?id=284541.

Fig. 1. A part of bug report for bug 284541 in Eclipse. This bug is about a missing node of XML files in Product WTP (Web Tools Platform). After the handling process, this bug is resolved as a fixed one.

TABLE 1
PART OF HISTORY OF BUG 284541 IN ECLIPSE

| Triager | Date | Action |
|---|---|---|
| Kaloyan Raev | 2009-08-12 | Assigned to the developer Kiril Mitov |
| Kaloyan Raev | 2010-01-14 | Assigned to the developer Kaloyan Raev |
| Kaloyan Raev | 2010-03-30 | Assigned to the developer Dimitar Giormov |
| Dimitar Giormov | 2010-04-12 | Changed status to assigned |
| Dimitar Giormov | 2010-04-14 | Changed status to resolved |
| | | Changed resolution to fixed |

A developer, who is assigned to a new bug report, starts to fix the bug based on the knowledge of historical bug fixing [64], [36]. Typically, the developer pays efforts to understand the new bug report and to examine historically fixed bugs as a reference (e.g., searching for similar bugs [54] and applying existing solutions to the new bug [28]).

An item status of a bug report is changed according to the current result of handling this bug until the bug is completely fixed. Changes of a bug report are stored in an item history. Table 1 presents a part of history of bug 284541. This bug has been assigned to three developers and only the last developer can handle this bug correctly. Changing developers lasts for over seven months while fixing this bug only costs three days.

Manual bug triage by a human triager is time-consuming and error-prone since the number of daily bugs is large to correctly assign and a human triager is hard to master the knowledge about all the bugs [12]. Existing work employs the approaches based on text classification to assist bug triage, e.g., [1], [25], [56]. In such approaches, the summary and the description of a bug report are extracted as the textual content while the developer who can fix this bug is marked as the label for classification. Then techniques on text classification can be used to predict the developer for a new bug. In details, existing bug reports with their developers are formed as a training set to train a classifier (e.g., Naive Bayes, a typical classifier in bug triage [12], [1], [25]); new bug reports are treated as a test set to examine the results of the classification. In Fig. 2(a), we illustrate the basic framework of bug triage based on text classification. As shown in Fig. 2(a), we view a bug data set as a text matrix. Each row of the matrix indicates one bug report while each column of the matrix indicates one word. To avoid the low accuracy of bug triage, a recommendation list with the size $k$ is used to provide a list of $k$ developers, who have the top-$k$ possibility to fix the new bug.

## 2.2 Motivation

Real-world data always include noise and redundancy [31]. Noisy data may mislead the data analysis techniques [66] while redundant data may increase the cost of data processing [19]. In bug repositories, all the bug reports are filled by developers in natural languages. The low-quality bugs accumulate in bug repositories with the growth in scale. Such large-scale and low-quality bug data may deteriorate the effectiveness of fixing bugs [28], [64]. In the following of this section, we will employ three examples of bug reports in Eclipse to show the motivation of our work, i.e., the necessity for data reduction.

We list the bug report of bug 205900 of Eclipse in Example 1 (the description in the bug report is partially omitted) to study the words of bug reports.

**Example 1**. *(Bug 205900) Current version in Eclipse Europa discovery repository broken.*

*... [Plug-ins] all installed correctly and do not show any errors in Plug-in configuration view. Whenever I try to add a [diagram name] diagram, the wizard cannot be started due to a missing [class name] class ...*

In this bug report, some words, e.g., installed, show, started, and missing, are commonly used for describing bugs. For text classification, such common words are not helpful for the quality of prediction. Hence, we tend to remove these words to reduce the computation for bug triage. However, for the text classification, the redundant words in bugs cannot be removed directly. Thus, we want to adapt a relevant technique for bug triage.

To study the noisy bug report, we take the bug report of bug 201598 as Example 2 (Note that both the summary and the description are included).

**Example 2**. *(Bug 201598) 3.3.1 about says 3.3.0.*
*Build id: M20070829-0800. 3.3.1 about says 3.3.0.*

This bug report presents the error in the version dialog. But the details are not clear. Unless a developer is very familiar with the background of this bug, it is hard to find the details. According to the item history, this bug is fixed by the developer who has reported this bug. But the summary of this bug may make other developers confused. Moreover, from the perspective of data processing, especially automatic processing, the words in this bug may be removed since these words are not helpful to identify this bug. Thus, it is necessary to remove the noisy bug reports and words for bug triage.

To study the redundancy between bug reports, we list two bug reports of bugs 200019 and 204653 in Example 3 (the items description are omitted).

**Example 3**. *Bugs 200019 and 204653.*
*(Bug 200019) Argument popup not highlighting the correct argument ...*
*(Bug 204653) Argument highlighting incorrect ...*

In bug repositories, the bug report of bug 200019 is

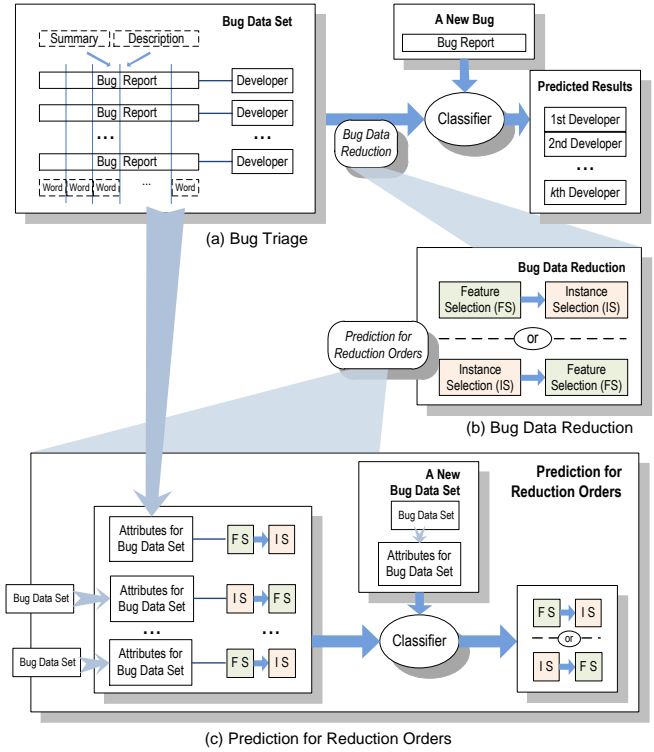

Fig. 2. Illustration of reducing bug data for bug triage. Sub-figure (a) presents the framework of existing work on bug triage. Before training a classifier with a bug data set, we add a phase of data reduction, in (b), which combines the techniques of instance selection and feature selection to reduce the scale of bug data. In bug data reduction, a problem is how to determine the order of two reduction techniques. In (c), based on the attributes of historical bug data sets, we propose a binary classification method to predict reduction orders.

marked as a duplicate one of bug 204653 (a *duplicate bug report*, denotes that a bug report describes one software fault, which has the same root cause as an existing bug report [54]). The textual contents of these two bug reports are similar. Hence, one of these two bug reports may be chosen as the representative one. Thus, we want to use a certain technique to remove one of these bug reports. Thus, a technique to remove extra bug reports for bug triage is needed.

Based on the above three examples, it is necessary to propose an approach to reducing the scale (e.g., large scale words in Example 1) and augmenting the quality of bug data (e.g., noisy bug reports in Example 2 and redundant bug reports in Example 3).

## 3 DATA REDUCTION FOR BUG TRIAGE

Motivated by the three examples in Section 2.2, we propose bug data reduction to reduce the scale and to improve the quality of data in bug repositories.

Fig. 2 illustrates the bug data reduction in our work, which is applied as a phase in data preparation of bug triage. We combine existing techniques of instance selection and feature selection to remove certain bug reports and words, i.e., in Fig. 2(b). A problem for reducing the bug data is to determine the order of applying instance selection and feature selection, which is denoted as the prediction of reduction orders, i.e., in Fig. 2(c).

In this section, we first present how to apply instance selection and feature selection to bug data, i.e., data reduction for bug triage. Then, we list the benefit of the data reduction. The details of the prediction for reduction orders will be shown in Section 4.

### 3.1 Applying Instance Selection and Feature Selection

In bug triage, a bug data set is converted into a text matrix with two dimensions, namely the bug dimension and the word dimension. In our work, we leverage the combination of instance selection and feature selection to generate a reduced bug data set. We replace the original data set with the reduced data set for bug triage.

Instance selection and feature selection are widely used techniques in data processing. For a given data set in a certain application, instance selection is to obtain a subset of relevant instances (i.e., bug reports in bug data) [18] while feature selection aims to obtain a subset of relevant features (i.e., words in bug data) [19]. In our work, we employ the combination of instance selection and feature selection. To distinguish the orders of applying instance selection and feature selection, we give the following denotation. Given an instance selection algorithm IS and a feature selection algorithm FS, we use FS→IS to denote the bug data reduction, which first applies FS and then IS; on the other hand, IS→FS denotes first applying IS and then FS.

In Algorithm 1, we briefly present how to reduce the bug data based on FS→IS. Given a bug data set, the output of bug data reduction is a new and reduced data set. Two algorithms FS and IS are applied sequentially. Note that in Step 2), some of bug reports may be blank during feature selection, i.e., all the words in a bug report are removed. Such blank bug reports are also removed in the feature selection.

In our work, FS→IS and IS→FS are viewed as two orders of bug data reduction. To avoid the bias from a single algorithm, we examine results of four typical algorithms of instance selection and feature selection, respectively. We briefly introduce these algorithms as follows.

Instance selection is a technique to reduce the number of instances by removing noisy and redundant instances [48], [11]. An instance selection algorithm can provide a reduced data set by removing non-representative instances [65], [38]. According to an existing comparison study [20] and an existing review [37], we choose four instance

| **Algorithm 1**. Data reduction based on FS→IS |  |
|---|---|
| **Input:** | training set $\mathcal{T}$ with $n$ words and $m$ bug reports, reduction order FS→IS<br>final number $n_F$ of words,<br>final number $m_I$ of bug reports, |
| **Output:** | reduced data set $\mathcal{T}_{FI}$ for bug triage |

1) apply FS to $n$ words of $\mathcal{T}$ and calculate objective values for all the words;
2) select the top $n_F$ words of $\mathcal{T}$ and generate a training set $\mathcal{T}_F$;
3) apply IS to $m_I$ bug reports of $\mathcal{T}_F$;
4) terminate IS when the number of bug reports is equal to or less than $m_I$ and generate the final training set $\mathcal{T}_{FI}$.

selection algorithms, namely Iterative Case Filter (ICF) [8], Learning Vectors Quantization (LVQ) [27], Decremental Reduction Optimization Procedure (DROP) [52], and Patterns by Ordered Projections (POP) [41].

Feature selection is a preprocessing technique for selecting a reduced set of features for large-scale data sets [19], [15]. The reduced set is considered as the representative features of the original feature set [10]. Since bug triage is converted into text classification, we focus on the feature selection algorithms in text data. In this paper, we choose four well-performed algorithms in text data [60], [43] and software data [49], namely Information Gain (IG) [24], $\chi^2$ statistic (CH) [60], Symmetrical Uncertainty attribute evaluation (SU) [51], and Relief-F Attribute selection (RF) [42]. Based on feature selection, words in bug reports are sorted according to their feature values and a given number of words with large values are selected as representative features.

### 3.2 Benefit of Data Reduction

In our work, to save the labor cost of developers, the data reduction for bug triage has two goals, 1) reducing the data scale and 2) improving the accuracy of bug triage. In contrast to modeling the textual content of bug reports in existing work (e.g., [1], [12], [25]), we aim to augment the data set to build a preprocessing approach, which can be applied before an existing bug triage approach. We explain the two goals of data reduction as follows.

1) **Reducing the data scale**. We reduce scales of data sets to save the labor cost of developers.

*Bug dimension*. As mentioned in Section 2.1, the aim of bug triage is to assign developers for bug fixing. Once a developer is assigned to a new bug report, the developer can examine historically fixed bugs to form a solution to the current bug report [64], [36]. For example, historical bugs are checked to detect whether the new bug is the duplicate of an existing one [54]; moreover, existing solutions to bugs can be searched and applied to the new bug [28]. Thus, we consider reducing duplicate and noisy bug reports to decrease the number of historical bugs. In practice, the labor cost of developers (i.e., the cost of examining historical bugs) can be saved by decreasing the number of bugs based on instance selection.

*Word dimension*. We use feature selection to remove noisy or duplicate words in a data set. Based on feature selection, the reduced data set can be handled more easily by automatic techniques (e.g., bug triage approaches) than the original data set. Besides bug triage, the reduced data set can be further used for other software tasks after bug triage (e.g., severity identification, time prediction, and reopened-bug analysis in Section 7.2).

2) **Improving the accuracy**. Accuracy is an important evaluation criterion for bug triage. In our work, data reduction explores and removes noisy or duplicate information in data sets (see examples in Section 2.2).

*Bug dimension*. Instance selection can remove uninformative bug reports; meanwhile, we can observe that the accuracy may be decreased by removing bug reports (see experiments in Section 5.2.3).

*Word dimension*. By removing uninformative words,

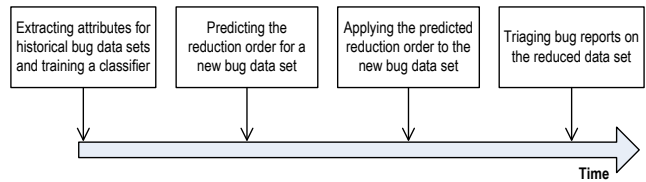

Fig. 3. Steps of predicting reduction orders for bug triage.

feature selection improves the accuracy of bug triage (see experiments in Section 5.2.3). This can recover the accuracy loss by instance selection.

## 4 PREDICTION FOR REDUCTION ORDERS

Based on Section 3.1, given an instance selection algorithm IS and a feature selection algorithm FS, FS→IS and IS→FS are viewed as two orders for applying reducing techniques. Hence, a challenge is how to determine the order of reduction techniques, i.e., how to choose one between FS→IS and IS→FS. We refer to this problem as the *prediction for reduction orders*.

### 4.1 Reduction Orders

To apply the data reduction to each new bug data set, we need to check the accuracy of both two orders (FS→IS and IS→FS) and choose a better one. To avoid the time cost of manually checking both reduction orders, we consider predicting the reduction order for a new bug data set based on historical data sets.

As shown in Fig. 2(c), we convert the problem of prediction for reduction orders into a binary classification problem. A bug data set is mapped to an instance and the associated reduction order (either FS→IS or IS→FS) is mapped to the label of a class of instances. Fig. 3 summarizes the steps of predicting reduction orders for bug triage. Note that a classifier can be trained only once when facing many new bug data sets. That is, training such a classifier once can predict the reduction orders for all the new data sets without checking both reduction orders. To date, the problem of predicting reduction orders of applying feature selection and instance selection has not been investigated in other application scenarios.

From the perspective of software engineering, predicting the reduction order for bug data sets can be viewed as a kind of *software metrics*, which involves activities for measuring some property for a piece of software [16]. However, the features in our work are extracted from the bug data set while the features in existing work on software metrics are for individual software artifacts[3], e.g., an individual bug report or an individual piece of code. In this paper, to avoid ambiguous denotations, an *attribute* refers to an extracted feature of a bug data set while a *feature* refers to a word of a bug report.

### 4.2 Attributes for a Bug Data Set

To build a binary classifier to predict reduction orders, we extract 18 attributes to describe each bug data set. Such attributes can be extracted before new bugs are triaged.

---

[3] In software metrics, a software artifact is one of many kinds of tangible products produced during the development of software, e.g., a use case, requirements specification, and a design document [16].

TABLE 2
AN OVERVIEW OF ATTRIBUTES FOR A BUG DATA SET

| Index | Attribute name | Description |
|---|---|---|
| B1 | # Bug reports | Total number of bug reports. |
| B2 | # Words | Total number of words in all the bug reports. |
| B3 | Length of bug reports | Average number of words of all the bug reports. |
| B4 | # Unique words | Average number of unique words in each bug report. |
| B5 | Ratio of sparseness | Ratio of sparse terms in the text matrix. A sparse term refers to a word with zero frequency in the text matrix. |
| B6 | Entropy of severities | Entropy of severities in bug reports. Severity denotes the importance of bug reports. |
| B7 | Entropy of priorities | Entropy of priorities in bug reports. Priority denotes the level of bug reports. |
| B8 | Entropy of products | Entropy of products in bug reports. Product denotes the sub-project. |
| B9 | Entropy of components | Entropy of components in bug reports. Component denotes the sub-sub-project. |
| B10 | Entropy of words | Entropy of words in bug reports. |
| D1 | # Fixers | Total number of developers who will fix bugs. |
| D2 | # Bug reports per fixer | Average number of bug reports for each fixer |
| D3 | # Words per fixer | Average number of words for each fixer |
| D4 | # Reporters | Total number of developers who have reported bugs. |
| D5 | # Bug reports per reporter | Average number of bug reports for each reporter |
| D6 | # Words per reporter | Average number of words for each reporter |
| D7 | # Bug reports by top 10% reporters | Ratio of bugs, which are reported by the most active reporters. |
| D8 | Similarity between fixers and reporters | Similarity between the set of fixers and the set of reporters, defined as the Tanimoto similarity. |

We divide these 18 attributes into two categories, namely the bug report category (B1 to B10) and the developer category (D1 to D8).

In Table 2, we present an overview of all the attributes of a bug data set. Given a bug data set, all these attributes are extracted to measure the characteristics of the bug data set. Among the attributes in Table 2, four attributes are directly counted from a bug data set, i.e., B1, B2, D1, and D4; six attributes are calculated based on the words in the bug data set, i.e., B3, B4, D2, D3, D5, and D6; five attributes are calculated as the entropy of an enumeration value to indicate the distributions of items in bug reports, i.e., B6, B7, B8, B9, and B10; three attributes are calculated according to the further statistics, i.e., B5, D7, and D8. All the 18 attributes in Table 2 can be obtained by direct extraction or automatic calculation. Details of calculating these attributes can be found in Section S2 in the supplemental material.

## 5 EXPERIMENTS AND RESULTS

### 5.1 Data Preparation

In this part, we present the data preparation for applying the bug data reduction. We evaluate the bug data reduction on bug repositories of two large open source projects, namely Eclipse and Mozilla. Eclipse [13] is a multi-language software development environment, including an Integrated Development Environment (IDE) and an extensible plug-in system; Mozilla [33] is an Internet application suite, including some classic products, such as the Firefox browser and the Thunderbird email client. Up to Dec. 31, 2011, 366,443 bug reports over 10 years have been recorded to Eclipse while 643,615 bug reports over 12 years have been recorded to Mozilla. In our work, we collect continuous 300,000 bug reports for each project of Eclipse and Mozilla, i.e., bugs 1-300000 in Eclipse and bugs 300001-600000 in Mozilla. Actually, 298,785 bug reports in Eclipse and 281,180 bug reports in Mozilla are collected since some of bug reports are removed from bug repositories (e.g., bug 5315 in Eclipse) or not allowed anonymous access (e.g., bug 40020 in Mozilla). For each bug report, we download web pages from bug repositories and extract the details of bug reports for experiments.

Since bug triage aims to predict the developers who can fix the bugs, we follow the existing work [1], [34] to remove unfixed bug reports, e.g., the new bug reports or will-not-fix bug reports. Thus, we only choose bug reports, which are fixed and duplicate (based on the items status of bug reports). Moreover, in bug repositories, several developers have only fixed very few bugs. Such inactive developers may not provide sufficient information for predicting correct developers. In our work, we remove the developers, who have fixed less than 10 bugs.

To conduct text classification, we extract the summary and the description of each bug report to denote the content of the bug. For a newly reported bug, the summary and the description are the most representative items, which are also used in manual bug triage [1]. As the input of classifiers, the summary and the description are converted into the vector space model [59], [4]. We employ two steps to form the word vector space, namely tokenization and stop word removal. First, we tokenize the summary and the description of bug reports into word vectors. Each word in a bug report is associated with its word frequency, i.e., the times that this word appears in the bug. Non-alphabetic words are removed to avoid the noisy words, e.g., memory address like 0x0902f00 in bug 200220 of Eclipse. Second, we remove the stop words, which are in high frequency and provide no helpful information for bug triage, e.g., the word "the" or "about". The list of stop words in our work is according to SMART information retrieval system [59]. We do not use the stemming technique in our work since existing work [12], [1] has examined that the stemming technique is not helpful to bug triage. Hence, the bug reports are converted into vector space model for further experiments.

### 5.2 Experiments on Bug Data Reduction

#### 5.2.1 Data sets and evaluation

We examine the results of bug data reduction on bug repositories of two projects, Eclipse and Mozilla. For each project, we evaluate results on five data sets and each data set is over 10,000 bug reports, which are fixed or duplicate bug reports. We check bug reports in the two projects and find out that 45.44% of bug reports in Eclipse and 28.23% of bug reports in Mozilla are fixed or duplicate. Thus, to obtain over 10,000 fixed or duplicate bug reports, each data set in Eclipse is collected from continuous 20,000 bug reports while each bug set in Mozilla is collected from continuous 40,000 bug reports. Table 3 lists

TABLE 3
TEN DATA SETS IN ECLIPSE AND MOZILLA

| | Name | DS-E1 | DS-E2 | DS-E3 | DS-E4 | DS-E5 |
|---|---|---|---|---|---|---|
| Eclipse | Range of Bug IDs | 200001 - 220000 | 220001 - 240000 | 240001 - 260000 | 260001 - 280000 | 280001 - 300000 |
| | # Bug reports | 11,313 | 11,788 | 11,495 | 11,401 | 10,404 |
| | # Words | 38,650 | 39,495 | 38,743 | 38,772 | 39,333 |
| | # Developers | 266 | 266 | 286 | 260 | 256 |
| | Name | DS-M1 | DS-M2 | DS-M3 | DS-M4 | DS-M5 |
| Mozilla | Range of Bug IDs | 400001 - 440000 | 440001 - 480000 | 480001 - 520000 | 520001 - 560000 | 560001 - 600000 |
| | # Bug reports | 14,659 | 14,746 | 16,479 | 15,483 | 17,501 |
| | # Words | 39,749 | 39,113 | 39,610 | 40,148 | 41,577 |
| | # Developers | 202 | 211 | 239 | 242 | 273 |

the details of ten data sets after data preparation.

To examine the results of data reduction, we employ four instance selection algorithms (ICF, LVQ, DROP, and POP), four feature selection algorithms (IG, CH, SU, and RF), and three bug triage algorithms (Support Vector Machine, SVM; K-Nearest Neighbor, KNN; and Naive Bayes, which are typical text-based algorithms in existing work [1], [25], [3]). Fig. 4 summarizes these algorithms. The implementation details can be found in Section S3 in the supplemental material.

The results of data reduction for bug triage can be measured in two aspects, namely the scales of data sets and the quality of bug triage. Based on Algorithm 1, the scales of data sets (including the number of bug reports and the number of words) are configured as input parameters. The quality of bug triage can be measured with the accuracy of bug triage, which is defined as $Accuracy_k = \frac{\text{\# correctly assigned bug reports in } k \text{ candidates}}{\text{\# all bug reports in the test set}}$. For each data set in Table 3, the first 80% of bug reports are used as a training set and the left 20% of bug reports are as a test set. In the following of this paper, data reduction on a data set is used to denote the data reduction on the training set of this data set since we cannot change the test set.

### 5.2.2 Rates of selected bug reports and words

For either instance selection or feature selection algorithm, the number of instances or features should be determined to obtain the final scales of data sets. We investigate the changes of accuracy of bug triage by varying the rate of selected bug reports in instance selection and the rate of selected words in feature selection. Taking two instance selection algorithms (ICF and LVQ) and two feature selection algorithms (IG and CH) as examples, we evaluate results on two data sets (DS-E1 in Eclipse and DS-M1 in Mozilla). Fig. 5 presents the accuracy of instance selection and feature selection (each value is an average of ten independent runs) for a bug triage algorithm, Naive Bayes.

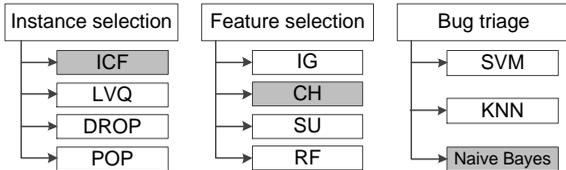

Fig. 4. Algorithms for instance selection, feature selection, and bug triage. Among these algorithms, ICF, CH, and Naive Bayes are well-performed based on the experiments of the bug data reduction.

For instance selection, ICF is a little better than LVQ from Fig. 5(a) and Fig. 5(c). A good percentage of bug reports is 50% or 70%. For feature selection, CH always performs better than IG from Fig. 5(b) and Fig. 5(d). We can find that 30% or 50% is a good percentage of words. In the other experiments, we directly set the percentages of selected bug reports and words to 50% and 30%, respectively.

### 5.2.3 Results of data reduction for bug triage

We evaluate the results of data reduction for bug triage on data sets in Table 3. First, we individually examine each instance selection algorithm and each feature selection algorithm based on one bug triage algorithm, Naive Bayes. Second, we combine the best instance selection algorithm and the best feature selection algorithm to examine the data reduction on three text-based bug triage algorithms.

In Tables 4 to 7, we show the results of four instance selection algorithms and four feature selection algorithms on four data sets in Table 3, i.e. DS-E1, DS-E5, DS-M1, and DS-M5. The best results by instance selection and the best results by feature selection are shown in bold. Results by Naive Bayes without instance selection or feature selection are also presented for comparison. The size of the recommendation list is set from 1 to 5. Results of the other six data sets in Table 3 can be found in Section S5 in the supplemental material. Based on Section 5.2.2, given a data set, IS denotes the 50% of bug reports are selected

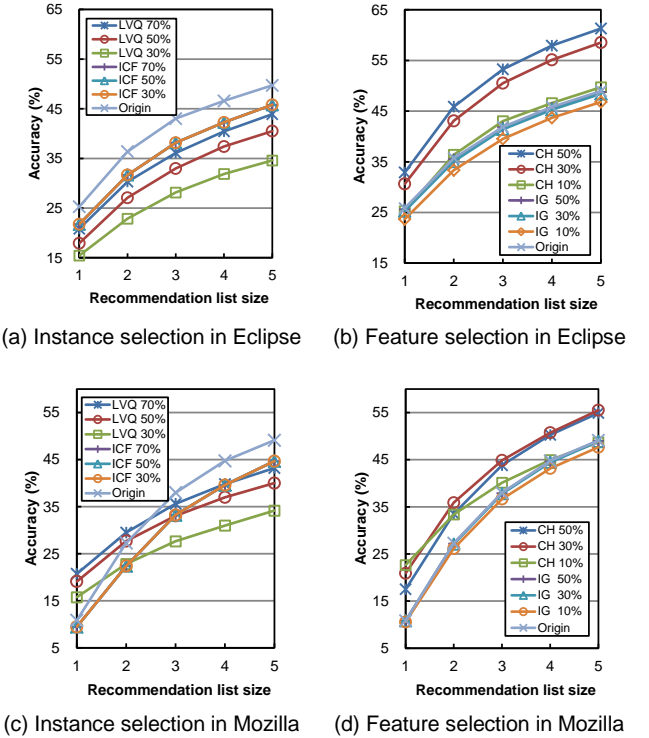

Fig. 5. Accuracy for instance selection or feature selection on Eclipse (DS-E1) and Mozilla (DS-M1). For instance selection, 30%, 50%, and 70% of bug reports are selected while for feature selection, 10%, 30%, and 50% of words are selected. The origin denotes the results of Naive Bayes without instance selection or feature selection. Note that some curves of ICF may be overlapped since ICF cannot precisely set the rate of final instances [8].

TABLE 4
ACCURACY (%) OF IS AND FS ON DS-E1

| List size | Origin | IS | | | | FS | | | |
|---|---|---|---|---|---|---|---|---|---|
| | | ICF | LVQ | DROP | POP | IG | CH | SU | RF |
| 1 | 25.85 | 21.75 | 17.91 | **22.53** | 20.36 | 25.27 | **30.64** | 23.64 | 24.52 |
| 2 | 35.71 | **31.66** | 27.08 | 31.40 | 29.59 | 35.07 | **43.09** | 33.44 | 34.87 |
| 3 | 41.88 | **38.17** | 32.97 | 36.64 | 36.01 | 41.42 | **50.52** | 40.18 | 40.93 |
| 4 | 45.84 | **42.25** | 37.40 | 40.10 | 40.45 | 45.26 | **55.12** | 44.90 | 45.01 |
| 5 | 48.95 | **45.79** | 40.50 | 42.76 | 44.16 | 48.42 | **58.54** | 47.95 | 47.90 |

TABLE 5
ACCURACY (%) OF IS AND FS ON DS-E5

| List size | Origin | IS | | | | FS | | | |
|---|---|---|---|---|---|---|---|---|---|
| | | ICF | LVQ | DROP | POP | IG | CH | SU | RF |
| 1 | 23.58 | 19.60 | 18.85 | 18.38 | **19.66** | 22.92 | **32.71** | 24.55 | 21.81 |
| 2 | 31.94 | **28.23** | 26.24 | 25.24 | 27.26 | 31.35 | **44.97** | 34.30 | 30.45 |
| 3 | 37.02 | **33.64** | 31.17 | 29.85 | 31.11 | 36.35 | **51.73** | 39.93 | 35.80 |
| 4 | 40.94 | **37.58** | 34.78 | 33.56 | 36.28 | 40.25 | **56.58** | 44.20 | 39.70 |
| 5 | 44.11 | **40.87** | 37.72 | 37.02 | 39.91 | 43.40 | **60.40** | 47.76 | 42.99 |

TABLE 6
ACCURACY (%) OF IS AND FS ON DS-M1

| List size | Origin | IS | | | | FS | | | |
|---|---|---|---|---|---|---|---|---|---|
| | | ICF | LVQ | DROP | POP | IG | CH | SU | RF |
| 1 | 10.86 | 9.46 | 19.10 | 11.06 | **21.07** | 10.80 | **20.91** | 17.53 | 11.01 |
| 2 | 27.29 | 22.39 | 27.70 | 27.77 | **29.13** | 27.08 | **35.88** | 30.37 | 27.26 |
| 3 | 37.99 | 33.23 | 33.06 | **36.33** | 32.81 | 37.77 | **44.86** | 38.66 | 37.27 |
| 4 | 44.74 | 39.60 | 36.99 | **41.77** | 38.82 | 44.43 | **50.73** | 44.35 | 43.95 |
| 5 | 49.11 | **44.68** | 40.01 | 44.56 | 42.68 | 48.87 | **55.50** | 48.36 | 48.33 |

TABLE 7
ACCURACY (%) OF IS AND FS ON DS-M5

| List size | Origin | IS | | | | FS | | | |
|---|---|---|---|---|---|---|---|---|---|
| | | ICF | LVQ | DROP | POP | IG | CH | SU | RF |
| 1 | 20.72 | 18.84 | **20.78** | 19.76 | 19.73 | 20.57 | **21.61** | 20.07 | 20.16 |
| 2 | 30.37 | 27.36 | 29.10 | 28.39 | **29.52** | 30.14 | **32.43** | 30.37 | 29.30 |
| 3 | 35.53 | 32.66 | 34.76 | 33.00 | **35.80** | 35.31 | **38.88** | 36.56 | 34.59 |
| 4 | 39.48 | 36.82 | 38.82 | 36.42 | **40.44** | 39.17 | **43.14** | 41.28 | 38.72 |
| 5 | 42.61 | 40.18 | 41.94 | 39.71 | **44.13** | 42.35 | **46.46** | 44.75 | 42.07 |

and FS denotes the 30% of words are selected.

As shown in Tables 4 and 5 for data sets in Eclipse, ICF provides eight best results among four instance selection algorithms when the list size is over two while either DROP or POP can achieve one best result when the list size is one. Among four feature selection algorithms, CH provides the best accuracy. IG and SU also achieve good results. In Tables 6 and 7 for Mozilla, POP in instance selection obtains six best results; ICF, LVQ, and DROP obtain one, one, two best results, respectively. In feature selection, CH also provides the best accuracy. Based on Tables 4 to 7, in the following of this paper, we only investigate the results of ICF and CH and to avoid the exhaustive comparison on all the four instance selection algorithms and four feature selection algorithms.

As shown in Tables 4 to 7, feature selection can increase the accuracy of bug triage over a data set while instance selection may decrease the accuracy. Such an accuracy decrease is coincident with existing work ([8], [20], [52], [41]) on typical instance selection algorithms on classic data sets[4], which shows that instance selection may hurt the accuracy. In the following, we will show that the accuracy decrease by instance selection is caused by the large number of developers in bug data sets.

To investigate the accuracy decrease by instance selection, we define the loss from origin to ICF as $Loss_k = \frac{Accuracy_k\ by\ origin - Accuracy_k\ by\ ICF}{Accuracy_k\ by\ origin}$, where the recommendation list size is $k$. Given a bug data set, we sort developers by the number of their fixed bugs in descending order. That is, we sort classes by the number of instances in classes. Then a new data set with $s$ developers is built by selecting the top-$s$ developers. For one bug data set, we build new data sets by varying $s$ from 2 to 30. Fig. 6 presents the loss on two bug data sets (DS-E1 and DS-M1) when $k = 1$ or $k = 3$.

As shown in Fig. 6, most of the loss from origin to ICF increases with the number of developers in the data sets. In other words, the large number of classes causes the accuracy decrease. Let us recall the data scales in Table 3. Each data set in our work contains over 200 classes. When applying instance selection, the accuracy of bug data sets in Table 3 may decrease more than that of the classic data sets in [8], [20], [52], [41] (which contain less than 20 classes and mostly 2 classes).

[4] UCI Machine Learning Repository, http://archive.ics.uci.edu/ml/.

In our work, the accuracy increase by feature selection and the accuracy decrease by instance selection lead to the combination of instance selection and feature selection. In other words, feature selection can supplement the loss of accuracy by instance selection. Thus, we apply instance selection and feature selection to simultaneously reduce the data scales. Tables 8 to 11 show the combinations of CH and ICF based on three bug triage algorithms, namely SVM, KNN, and Naive Bayes, on four data sets.

As shown in Table 8, for the Eclipse data set DS-E1, ICF→CH provides the best accuracy on three bug triage algorithms. Among these algorithms, Naive Bayes can obtain much better results than SVM and KNN. ICF→CH based on Naive Bayes obtains the best results. Moreover, CH→ICF based on Naive Bayes can also achieve good results, which are better than Naive Bayes without data reduction. Thus, data reduction can improve the accuracy of bug triage, especially, for the well-performed algorithm, Naive Bayes.

In Tables 9 to 11, data reduction can also improve the accuracy of KNN and Naive Bayes. Both CH→ICF and ICF→CH can obtain better solutions than the origin bug triage algorithms. An exceptional algorithm is SVM. The accuracy of data reduction on SVM is lower than that of the original SVM. A possible reason is that SVM is a kind of discriminative model, which is not suitable for data reduction and has a more complex structure than KNN and Naive Bayes.

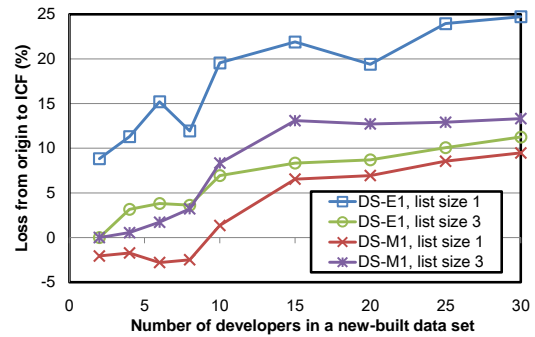

Fig. 6. Loss from origin to ICF on two data sets. The origin denotes the bug triage algorithm, Naive Bayes. The x-axis is the number of developers in a new-built data set; the y-axis is the loss. The loss above zero denotes the accuracy of ICF is lower than that of origin while the loss below zero denotes the accuracy of ICF is higher than that of origin.

TABLE 8
ACCURACY (%) OF DATA REDUCTION ON DS-E1

| List size | SVM | | | KNN | | | Naive Bayes | | |
|---|---|---|---|---|---|---|---|---|---|
| | Origin | CH→ICF | ICF→CH | Origin | CH→ICF | ICF→CH | Origin | CH→ICF | ICF→CH |
| 1 | 7.75 | 7.19 | **8.77** | 12.76 | 18.51 | **20.63** | 25.85 | 25.42 | **27.24** |
| 2 | 11.45 | 12.39 | **14.41** | 12.96 | 20.46 | **24.06** | 35.71 | 39.00 | **39.56** |
| 3 | 15.40 | 15.81 | **18.45** | 13.04 | 21.38 | **25.75** | 41.88 | 46.88 | **47.58** |
| 4 | 18.27 | 18.53 | **21.55** | 13.14 | 22.13 | **26.53** | 45.84 | 51.77 | **52.45** |
| 5 | 21.18 | 20.79 | **23.54** | 13.23 | 22.58 | **27.27** | 48.95 | 55.55 | **55.89** |

TABLE 9
ACCURACY (%) OF DATA REDUCTION ON DS-E5

| List size | SVM | | | KNN | | | Naive Bayes | | |
|---|---|---|---|---|---|---|---|---|---|
| | Origin | CH→ICF | ICF→CH | Origin | CH→ICF | ICF→CH | Origin | CH→ICF | ICF→CH |
| 1 | **6.21** | 5.05 | 5.83 | 14.78 | 19.11 | **22.81** | 23.58 | 27.93 | **28.81** |
| 2 | **10.18** | 7.77 | 8.99 | 15.09 | 21.21 | **25.85** | 31.94 | 40.16 | **40.44** |
| 3 | **12.87** | 10.27 | 11.19 | 15.34 | 22.21 | **27.29** | 37.02 | **47.92** | 47.19 |
| 4 | **16.21** | 12.19 | 13.12 | 15.45 | 22.85 | **28.13** | 40.94 | **52.91** | 52.18 |
| 5 | **18.14** | 14.18 | 14.97 | 15.55 | 23.21 | **28.61** | 44.11 | **56.25** | 55.51 |

TABLE 10
ACCURACY (%) OF DATA REDUCTION ON DS-M1

| List size | SVM | | | KNN | | | Naive Bayes | | |
|---|---|---|---|---|---|---|---|---|---|
| | Origin | CH→ICF | ICF→CH | Origin | CH→ICF | ICF→CH | Origin | CH→ICF | ICF→CH |
| 1 | **11.98** | 10.88 | 10.38 | 11.87 | 14.74 | **15.10** | 10.86 | 17.07 | **19.45** |
| 2 | **21.82** | 19.36 | 17.98 | 12.63 | 16.40 | **18.44** | 27.29 | 31.77 | **32.11** |
| 3 | **29.61** | 26.65 | 24.93 | 12.81 | 16.97 | **19.43** | 37.99 | **41.67** | 40.28 |
| 4 | **35.08** | 32.03 | 29.46 | 12.88 | 17.29 | **19.93** | 44.74 | **48.43** | 46.47 |
| 5 | **38.72** | 36.22 | 33.27 | 13.08 | 17.82 | **20.55** | 49.11 | **53.38** | 51.40 |

TABLE 11
ACCURACY (%) OF DATA REDUCTION ON DS-M5

| List size | SVM | | | KNN | | | Naive Bayes | | |
|---|---|---|---|---|---|---|---|---|---|
| | Origin | CH→ICF | ICF→CH | Origin | CH→ICF | ICF→CH | Origin | CH→ICF | ICF→CH |
| 1 | **15.01** | 14.87 | 14.24 | 13.92 | 14.66 | **16.66** | 20.72 | 20.97 | **21.88** |
| 2 | **21.64** | 20.45 | 20.10 | 14.75 | 16.62 | **18.85** | 30.37 | 31.27 | **32.91** |
| 3 | **25.65** | 24.26 | 23.82 | 14.91 | 17.70 | **19.84** | 35.53 | 37.24 | **39.70** |
| 4 | **28.36** | 27.18 | 27.21 | 15.36 | 18.37 | **20.78** | 39.48 | 41.59 | **44.50** |
| 5 | **30.73** | 29.51 | 29.79 | 15.92 | 19.07 | **21.46** | 42.61 | 45.28 | **48.28** |

As shown in Tables 8 to 11, all the best results are obtained by CH→ICF or ICF→CH based on Naive Bayes. Based on data reduction, the accuracy of Naive Bayes on Eclipse is improved by 2% to 12% and the accuracy on Mozilla is improved by 1% to 6% Considering the list size 5, data reduction based on Naive Bayes can obtain from 13% to 38% better results than that based on SVM and can obtain 21% to 28% better results than that based on KNN. We find out that data reduction should be built on a well-performed bug triage algorithm. In the following, we focus on the data reduction on Naive Bayes.

In Tables 8 to 11, the combinations of instance selection and feature selection can provide good accuracy and reduce the number of bug reports and words of the bug data. Meanwhile, the orders, CH→ICF and ICF→CH, lead to different results. Taking the list size five as an example, for Naive Bayes, CH→ICF provides better accuracy than ICF→CH on DS-M1 while ICF→CH provides better accuracy than CH→ICF on DS-M5.

In Table 12, we compare the time cost of data reduction with the time cost of manual bug triage on four data sets. As shown in Table 12, the time cost of manual bug triage is much longer than that of data reduction. For a bug report, the average time cost of manual bug triage is from 23 to 57 days. The average time of the original Naive Bayes is from 88 to 139 seconds while the average time of data reduction is from 298 to 1,558 seconds. Thus, compared with the manual bug triage, data reduction is efficient for bug triage and the time cost of data reduction can be ignored.

In summary of the results, data reduction for bug triage can improve the accuracy of bug triage to the original data set. The advantage of the combination of instance selection and feature selection is to improve the accuracy and to reduce the scales of data sets on both the bug dimension and the word dimension (removing 50% of bug reports and 70% of words).

### 5.2.4 A Brief Case Study

The results in Tables 8 to 11 show that the order of applying instance selection and feature selection can impact the final accuracy of bug triage. In this part, we employ ICF and CH with Naive Bayes to conduct a brief case study on the data set DS-E1.

First, we measure the differences of reduced data set by CH→ICF and ICF→CH. Fig. 7 illustrates bug reports and words in the data sets by applying CH→ICF and ICF→CH. Although there exists an overlap between the data sets by CH→ICF and ICF→CH, either CH→ICF or ICF→CH retains its own bug reports and words. For example, we can observe that the reduced data set by CH→ICF keeps 1,655 words, which have been removed by ICF→CH; the reduced data set by ICF→CH keeps 2,150 words, which have been removed by CH→ICF. Such observation indicates the orders of applying CH and ICF will brings different results for the reduced data set.

Second, we check the duplicate bug reports in the data sets by CH→ICF and ICF→CH. Duplicate bug reports are a kind of redundant data in a bug repository [54], [47]. Thus, we count the changes of duplicate bug reports in the data sets. In the original training set, there exist 532 duplicate bug reports. After data reduction, 198 duplicate bug reports are removed by CH→ICF while 262 are removed by ICF→CH. Such a result indicates that the order of applying instance selection and feature selection can impact the ability of removing redundant data.

Third, we check the blank bug reports during the data reduction. In this paper, a *blank bug report* refers to a zero-word bug report, whose words are removed by feature selection. Such blank bug reports are finally removed in the data reduction since they provides none of information. The removed bug reports and words can be viewed as a kind of noisy data. In our work, bugs 200019, 200632, 212996, and 214094 become blank bug reports after applying CH→ICF while bugs 201171, 201598, 204499, 209473, and 214035 become blank bug reports after ICF→CH. There is no overlap between the blank bug reports by CH→ICF and ICF→CH. Thus, we find out that the order of applying instance selection and feature selection also impacts the ability of removing noisy data.

TABLE 12
TIME COMPARISON BETWEEN DATA REDUCTION AND MANUAL WORK

| Data set | Manual bug triage | Origin | | | CH→ICF | | | | ICF→CH | | | |
|---|---|---|---|---|---|---|---|---|---|---|---|---|
| | | Preprocessing | Naive Bayes | Sum | Preprocessing | Data reduction | Naive Bayes | Sum | Preprocessing | Data reduction | Naive Bayes | Sum |
| DS-E1 | 32.55 day | 59 sec | 29 sec | 88 sec | 58 sec | 322 sec | 3 sec | 383 sec | 59 sec | 458 sec | 2 sec | 519 sec |
| DS-E5 | 23.14 day | 55 sec | 25 sec | 80 sec | 54 sec | 241 sec | 3 sec | 298 sec | 54 sec | 367 sec | 3 sec | 424 sec |
| DS-M1 | 57.44 day | 88 sec | 33 sec | 121 sec | 88 sec | 698 sec | 4 sec | 790 sec | 88 sec | 942 sec | 3 sec | 1,033 sec |
| DS-M5 | 23.77 day | 87 sec | 52 sec | 139 sec | 87 sec | 1,269 sec | 6 sec | 1,362 sec | 88 sec | 1,465 sec | 5 sec | 1,558 sec |

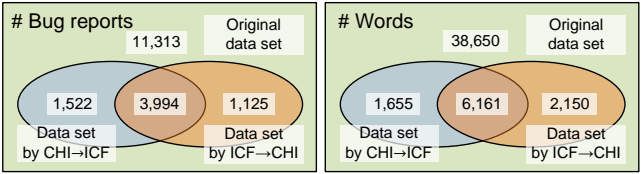

(a) Bug reports in data sets  (b) Words in data sets

Fig. 7. Bug reports and words in the data set DS-E1 (i.e., bugs 200001-220000 in Eclipse) by applying CH→ICF and ICF→CH.

TABLE 14
DATA SETS OF PREDICTION FOR REDUCTION ORDERS

| Project | # Data sets | # CH→ICF | # ICF→CH |
|---|---|---|---|
| Eclipse | 300 | 45 | 255 |
| Mozilla | 399 | 157 | 242 |
| Eclipse & Mozilla | 699 | 202 | 497 |

In summary of this brief case study on the data set in Eclipse, the results of data reduction are impacted by the order of applying instance selection and feature selection. Thus, it is necessary to investigate how to determine the order of applying these algorithms.

To further examine whether the results by CH→ICF are significantly different from those by ICF→CH, we perform a Wilcoxon signed-rank test [53] on the results by CH→ICF and ICF→CH on 10 data sets in Table 3. In details, we collect 50 pairs of accuracy values (10 data sets; 5 recommendation lists for each data set, i.e. the size from 1 to 5) by applying CH→ICF and ICF→CH, respectively. The result of test is with a statistically significant $p$-value of 0.018, i.e., applying CH→ICF or ICF→CH leads to significantly differences for the accuracy of bug triage.

## 5.3 Experiments on Prediction for Reduction Orders

### 5.3.1 Data sets and evaluation

We present the experiments on prediction for reduction orders in this part. We map a bug data set to an instance, and map the reduction order (i.e., FS→IS or IS→FS) to its label. Given a new bug data set, we train a classifier to predict its appropriate reduction order based on historical bug data sets.

As shown in Fig. 2(c), to train the classifier, we label each bug data set with its reduction order. In our work, one *bug unit* denotes 5,000 continuous bug reports. In Section 5.1, we have collected 298,785 bug reports in Eclipse and 281,180 bug reports in Mozilla. Then, 60 bug units

TABLE 13
SETUP OF DATA SETS IN ECLIPSE

| #Bug units | Data sets [a] |
|---|---|
| 1 | $\{u_1\}, \{u_2\}, \{u_3\}, \ldots, \{u_{60}\}$ |
| 2 | $\{u_1, u_2\}, \{u_2, u_3\}, \{u_3, u_4\}, \ldots, \{u_{60}, u_1\}$ |
| 3 | $\{u_1, u_2, u_3\}, \{u_2, u_3, u_4\}, \{u_3, u_4, u_5\}, \ldots, \{u_{60}, u_1, u_2\}$ |
| 4 | $\{u_1, u_2, u_3, u_4\}, \{u_2, u_3, u_4, u_5\}, \{u_3, u_4, u_5, u_6\}, \ldots, \{u_{60}, u_1, u_2, u_3\}$ |
| 5 | $\{u_1, u_2, u_3, u_4, u_5\}, \{u_2, u_3, u_4, u_5, u_6\}, \{u_3, u_4, u_5, u_6, u_7\}, \ldots, \{u_{60}, u_1, u_2, u_3, u_4\}$ |

[a] *Each $\{\cdot\}$ denotes a bug data set, where $u_t$ denotes a bug unit ($1 \leq t \leq 60$).*

( 298,785/5,000 = 59.78 ) for Eclipse and 57 bug units (281,180/5,000 = 56.24) for Mozilla are obtained. Next, we form bug data sets by combining bug units to training classifiers. In Table 13, we show the setup of data sets in Eclipse. Given 60 bug units in Eclipse, we consider continuous one to five bug units as one data set. In total, we collect 300 (60 × 5) bug data sets on Eclipse. Similarly, we consider continuous one to seven bug units as one data set on Mozilla and finally collect 399 (57 × 7) bug data sets. For each bug data set, we extract 18 attributes according to Table 2 and normalize all the attributes to values between 0 and 1.

We examine the results of prediction of reduction orders on ICF and CH. Given ICF and CH, we label each bug data set with its reduction order (i.e., CH→ICF or ICF→CH). First, for a bug data set, we respectively obtain the results of CH→ICF and ICF→CH by evaluating data reduction for bug triage based on Section 5.2. Second, for a recommendation list with size 1 to 5, we count the times of each reduction order when the reduction order obtain the better accuracy. That is, if CH→ICF can provide more times of the better accuracy, we label the bug data set with CH→ICF, and verse vice.

Table 14 presents the statistics of bug data sets of Eclipse and Mozilla. Note that the numbers of data sets with CH→ICF and ICF→CH are imbalance. In our work, we employ the classifier AdaBoost to predict reduction orders since AdaBoost is useful to classify imbalanced data and generates understandable results of classification [24].

In experiments, 10-fold cross-validation is used to evaluate the prediction for reduction orders. We employ four evaluation criteria, namely precision, recall, $F_1$-measure, and accuracy. To balance the precision and recall, the $F_1$-measure is defined as $F_1 = \frac{2 \times Recall \times Precision}{Recall + Precision}$. For a good classifier, $F_{1\,CH \to ICF}$ and $F_{1\,ICF \to CH}$ should be balanced to avoid classifying all the data sets into only one class. The accuracy measures the percentage of correctly predicted orders over the total bug data sets. The accuracy is defined as $Accuracy = \frac{\#\ correctly\ predicted\ orders}{\#\ all\ data\ sets}$.

### 5.3.2 Results

We investigate the results of predicting reductions orders for bug triage on Eclipse and Mozilla. For each project, we employ AdaBoost as the classifier based on two strategies, namely resampling and reweighting [17]. A decision tree classifier, C4.5, is embedded into AdaBoost. Thus, we compare results of classifiers in Table 15.

In Table 15, C4.5, AdaBoost C4.5 resampling, and AdaBoost C4.5 reweighting, can obtain better values of $F_1$-measure on Eclipse and AdaBoost C4.5 reweighting obtains the best $F_1$-measure. All the three classifiers can obtain good accuracy and C4.5 can obtain the best accuracy. Due to the imbalanced number of bug data sets, the values of $F_1$-measure of CH→ICF and ICF→CH are imbalanced. The results on Eclipse indicate that AdaBoost with reweighting provides the best results among these three classifiers.

TABLE 15
RESULTS ON PREDICTING REDUCTION ORDERS (%)

| Project | Classifier | CH→ICF | | | ICF→CH | | | Accuracy |
|---|---|---|---|---|---|---|---|---|
| | | Precision | Recall | F₁ | Precision | Recall | F₁ | |
| Eclipse | C4.5 | 13.3 | 4.4 | 6.7 | 84.9 | 94.9 | **89.6** | **81.3** |
| | AdaBoost C4.5 resampling | 14.7 | 11.1 | 12.7 | 85.0 | 88.6 | 86.8 | 77.0 |
| | AdaBoost C4.5 reweighting | 16.7 | 15.6 | **16.1** | 85.3 | 86.3 | 85.8 | 75.7 |
| Mozilla | C4.5 | 48.0 | 29.9 | 36.9 | 63.5 | 78.9 | 70.3 | 59.6 |
| | AdaBoost C4.5 resampling | 52.7 | 56.1 | **54.3** | 70.3 | 67.4 | 68.8 | **62.9** |
| | AdaBoost C4.5 reweighting | 49.5 | 33.1 | 39.7 | 64.3 | 78.1 | **70.5** | 60.4 |

TABLE 17
TOP NODE ANALYSIS OF PREDICTING REDUCTION ORDERS

| Level [a] | Frequency | Index | Attribute name |
|---|---|---|---|
| 0 | 2 | B3 | Length of bug reports |
| | 2 | D3 | # Words per fixer |
| 1 | 3 | B6 | Entropy of severity |
| | 3 | D1 | # Fixers |
| | 2 | B3 | Length of bug reports |
| | 2 | B4 | # Unique words |
| 2 | 4 | B6 | Entropy of severity |
| | 3 | B7 | Entropy of priority |
| | 3 | B9 | Entropy of component |
| | 2 | B3 | Length of bug reports |
| | 2 | B4 | # Unique words |
| | 2 | B5 | Ratio of sparseness |
| | 2 | B8 | Entropy of product |
| | 2 | D5 | # Bug reports per reporter |
| | 2 | D8 | Similarity between fixers and reporters |

[a] Only nodes in Level 0 to Level 2 of decision trees are presented. In each level, we omit an attribute if its frequency equals to 1.

For the other project, Mozilla in Table 15, AdaBoost with resampling can obtain the best accuracy and F₁-measure. Note that the values of F₁-measure by CH→ICF and ICF→CH on Mozilla are more balanced than those on Eclipse. For example, when classifying with AdaBoost C4.5 reweighting, the difference of F₁-measure on Eclipse is 69.7% (85.8% − 16.1%) and the difference on Mozilla is 30.8% (70.5% − 39.7%). A reason for this fact is that the number of bug data sets with the order ICF→CH in Eclipse is about 5.67 times (255/45) of that with CH→ICF while in Mozilla, the number of bug data sets with ICF→CH is 1.54 times (242/157) of that with CH→ICF.

The number of bug data sets on either Eclipse (300 data sets) or Mozilla (399 data sets) is small. Since Eclipse and Mozilla are both large-scale open source projects and share the similar style in development [64], we consider combining the data sets of Eclipse and Mozilla to form a large amount of data sets. Table 16 shows the results of predicting reduction orders on totally 699 bug data sets, including 202 data sets with CH→ICF and 497 data sets with ICF→CH. As shown in Table 16, the results of three classifiers are very close. Each of C4.5, AdaBoost C4.5 resampling and AdaBoost C4.5 reweighting can provide good F₁-measure and accuracy. Based on the results of these 699 bug data sets in Table 16, AdaBoost C4.5 reweighting is the best one among these three classifiers.

As shown in Tables 15 and 16, we can find out that it is feasible to build a classifier based on attributes of bug data sets to determine using CH→ICF or ICF→CH. To investigate which attribute impacts the predicted results, we employ the *top node analysis* to further check the results by AdaBoost C4.5 reweighting in Table 16. Top node analysis is a method to rank representative nodes (e.g., attributes in prediction for reduction orders) in a decision tree classifier on software data [46].

In Table 17, we employ the top node analysis to present the representative attributes when predicting the reduction order. The level of a node denotes the distance to the root node in a decision tree (Level 0 is the root node); the frequency denotes the times of appearing in one level (the sum of ten decision trees in 10-fold cross-validation). In Level 0, i.e., the root node of decision trees, attributes B3 (Length of bug reports) and D3 (# Words per fixer) appear for two times. In other words, these two attributes are more decisive than the other attributes to predict the reduction orders. Similarly, B6, D1, B3, and B4 are decisive attributes in Level 1. By checking all the three levels in Table 17, the attribute B3 (Length of bug reports) appears in all the levels. This fact indicates that B3 is a representative attribute when predicting the reduction order. Moreover, based on the analysis in Table 17, no attribute dominates all the levels. For example, each attribute in Level 0 contributes to the frequency with no more than 2 and each attribute in Level 1 contributes to no more than 3. The results in the top node analysis indicate that only one attribute cannot determine the prediction of reduction orders and each attribute is helpful to the prediction.

TABLE 16
RESULTS ON PREDICTING REDUCTION ORDERS BY COMBINING BUG DATA SETS ON ECLIPSE AND MOZILLA (%)

| Classifier | CH→ICF | | | ICF→CH | | | Accuracy |
|---|---|---|---|---|---|---|---|
| | Precision | Recall | F₁ | Precision | Recall | F₁ | |
| C4.5 | 49.5 | 50.5 | **50.0** | 79.7 | 79.1 | 79.4 | 70.8 |
| AdaBoost C4.5 resampling | 49.4 | 40.1 | 44.3 | 77.4 | 83.3 | 80.2 | 70.8 |
| AdaBoost C4.5 reweighting | 51.3 | 48.0 | 49.6 | 79.4 | 81.5 | **80.4** | **71.8** |

## 6 DISCUSSION

In this paper, we propose the problem of data reduction for bug triage to reduce the scales of data sets and to improve the quality of bug reports. We use techniques of instance selection and feature selection to reduce noise and redundancy in bug data sets. However, not all the noise and redundancy are removed. For example, as mentioned in Section 5.2.4, only less than 50% of duplicate bug reports can be removed in data reduction (198/532 = 37.2% by CH→ICF and 262/532 = 49.2% by ICF→CH). The reason for this fact is that it is hard to exactly detect noise and redundancy in real-world applications. On one hand, due to the large scales of bug repositories, there exist no adequate labels to mark whether a bug report or a word belongs to noise or redundancy; on the other hand, since all the bug reports in a bug repository are recorded in natural languages, even noisy and redundant data may contain useful information for bug fixing.

In our work, we propose the data reduction for bug triage. As shown in Tables 4 to 7, although a recommendation list exists, the accuracy of bug triage is not good (less than 61%). This fact is caused by the complexity of bug triage. We explain such complexity as follows. First, in bug reports, statements in natural languages may be hard to clearly understand; second, there exist many potential developers in bug repositories (over 200 developers based on Table 3); third, it is hard to cover all the knowledge of bugs in a software project and even human triagers may assign developers by mistake. Our work can be used to assist human triagers rather than replace them.

In this paper, we construct a predictive model to determine the reduction order for a new bug data set based on historical bug data sets. Attributes in this model are statistic values of bug data sets, e.g., the number of words or the length of bug reports. No representative words of bug data sets are extracted as attributes. We plan to extract more detailed attributes in future work.

The values of $F_1$-measure and accuracy of prediction for reduction orders are not large enough for binary classifiers. In our work, we tend to present a resolution to determine the reduction order of applying instance selection and feature selection. Our work is not an ideal resolution to the prediction of reduction orders and can be viewed as a step towards the automatic prediction. We can train the predictive model once and predict reduction orders for each new bug data set. The cost of such prediction is not expensive, compared with trying all the orders for bug data sets.

Another potential issue is that bug reports are not reported at the same time in real-world bug repositories. In our work, we extract attributes of a bug data set and consider that all the bugs in this data set are reported in certain days. Compared with the time of bug triage, the time range of a bug data set can be ignored. Thus, the extraction of attributes from a bug data set can be applied to real-world applications.

## 7 RELATED WORK

In this section, we review existing work on modeling bug data, bug triage, and the quality of bug data with defect prediction.

### 7.1 Modeling Bug Data

To investigate the relationships in bug data, Sandusky et al. [45] form a bug report network to examine the dependency among bug reports. Besides studying relationships among bug reports, Hong et al. [23] build a developer social network to examine the collaboration among developers based on the bug data in Mozilla project. This developer social network is helpful to understand the developer community and the project evolution. By mapping bug priorities to developers, Xuan et al. [57] identify the developer prioritization in open source bug repositories. The developer prioritization can distinguish developers and assist tasks in software maintenance.

To investigate the quality of bug data, Zimmermann et al. [64] design questionnaires to developers and users in three open source projects. Based on the analysis of questionnaires, they characterize what makes a good bug report and train a classifier to identify whether the quality of a bug report should be improved. Duplicate bug reports weaken the quality of bug data by delaying the cost of handling bugs. To detect duplicate bug reports, Wang et al. [54] design a natural language processing approach by matching the execution information; Sun et al. [47] propose a duplicate bug detection approach by optimizing a retrieval function on multiple features.

To improve the quality of bug reports, Breu et al. [9] have manually analyzed 600 bug reports in open source projects to seek for ignored information in bug data. Based on the comparative analysis on the quality between bugs and requirements, Xuan et al. [55] transfer bug data to requirements databases to supplement the lack of open data in requirements engineering.

In this paper, we also focus on the quality of bug data. In contrast to existing work on studying the characteristics of data quality (e.g., [64], [9]) or focusing on duplicate bug reports (e.g., [54], [47]), our work can be utilized as a preprocessing technique for bug triage, which both improves data quality and reduces data scale.

### 7.2 Bug Triage

Bug triage aims to assign an appropriate developer to fix a new bug, i.e., to determine who should fix a bug. Čubranić & Murphy [12] first propose the problem of automatic bug triage to reduce the cost of manual bug triage. They apply text classification techniques to predict related developers. Anvik et al. [1] examine multiple techniques on bug triage, including data preparation and typical classifiers. Anvik & Murphy [3] extend above work to reduce the effort of bug triage by creating development-oriented recommenders.

Jeong et al. [25] find out that over 37% of bug reports have been reassigned in manual bug triage. They propose a tossing graph method to reduce reassignment in bug triage. To avoid low-quality bug reports in bug triage, Xuan et al. [56] train a semi-supervised classifier by combining unlabeled bug reports with labeled ones. Park et al. [40] convert bug triage into an optimization problem and propose a collaborative filtering approach to reducing the bug-fixing time.

For bug data, several other tasks exist once bugs are triaged. For example, severity identification [30] aims to detect the importance of bug reports for further scheduling in bug handling; time prediction of bugs [61] models the time cost of bug fixing and predicts the time cost of given bug reports; reopened-bug analysis [46], [63] identifies the incorrectly fixed bug reports to avoid delaying the software release.

In data mining, the problem of bug triage relates to the problems of *expert finding* (e.g., [6], [50]) and *ticket routing* (e.g., [44], [35]). In contrast to the broad domains in expert finding or ticket routing, bug triage only focuses on assign developers for bug reports. Moreover, bug reports in bug triage are transferred into documents (not keywords in expert finding) and bug triage is a kind of content-based classification (not sequence-based in ticket routing).

### 7.3 Data Quality in Defect Prediction

In our work, we address the problem of data reduction for bug triage. To our knowledge, no existing work has investigated the bug data sets for bug triage. In a related problem, *defect prediction*, some work has focused on the data quality of software defects. In contrast to multiple-class classification in bug triage, defect prediction is a binary-class classification problem, which aims to predict whether a software artifact (e.g., a source code file, a class, or a module) contains faults according to the extracted features of the artifact.

In software engineering, defect prediction is a kind of work on software metrics. To improve the data quality, Khoshgoftaar et al. [26] and Gao et al. [21] examine the techniques on feature selection to handle imbalanced defect data. Shivaji et al. [49] proposes a framework to examine multiple feature selection algorithms and remove noise features in classification-based defect prediction. Besides feature selection in defect prediction, Kim et al. [29] present how to measure the noise resistance in defect prediction and how to detect noise data. Moreover, Bishnu & Bhattacherjee [7] process the defect data with quad tree based k-means clustering to assist defect prediction.

In this paper, in contrast to the above work, we address the problem of data reduction for bug triage. Our work can be viewed as an extension of software metrics. In our work, we predict a value for a set of software artifacts while existing work in software metrics predict a value for an individual software artifact.

## 8 Conclusions

Bug triage is an expensive step of software maintenance in both labor cost and time cost. In this paper, we combine feature selection with instance selection to reduce the scale of bug data sets as well as improve the data quality. To determine the order of applying instance selection and feature selection for a new bug data set, we extract attributes of each bug data set and train a predictive model based on historical data sets. We empirically investigate the data reduction for bug triage in bug repositories of two large open source projects, namely Eclipse and Mozilla. Our work provides an approach to leveraging techniques on data processing to form reduced and high-quality bug data in software development and maintenance.

In future work, we plan on improving the results of data reduction in bug triage to explore how to prepare a high-quality bug data set and tackle a domain-specific software task. For predicting reduction orders, we plan to pay efforts to find out the potential relationship between the attributes of bug data sets and the reduction orders.

## Acknowledgment

The authors would like to thank the anonymous reviewers for their valuable and constructive comments on improving the paper. This work has been supported by the National Natural Science Foundation of China (under grants 61033012, 61229301, and 61370144), the New Century Excellent Talents in University (under grant NCET-13-0073), the Program for Changjiang Scholars and Innovative Research Team in University (PCSIRT) of the Ministry of Education, China (under grant IRT13059), and the National 973 Program of China (under grant 2013CB329604).

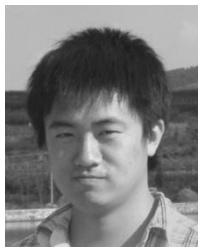

**Jifeng Xuan** received the BSc degree in software engineering in 2007 and the PhD degree in 2013, from Dalian University of Technology, China. He is currently a postdoctoral researcher of INRIA Lille – Nord Europe, France. His research interests include mining software repositories, search based software engineering, and machine learning. He is a member of the ACM and the China Computer Federation (CCF).

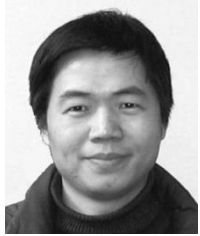

**He Jiang** is a professor of School of Software, Dalian University of Technology, China. His research interests include computational intelligence and its applications in software engineering and data mining. Jiang received his PhD degree in computer science from the University of Science and Technology of China (USTC), China. He is a program co-chair of the 2012 International Conference on Industrial, Engineering & Other Applications of Applied Intelligent Systems (IEA-AIE 2012). He is also a member of the IEEE, the ACM and the CCF.

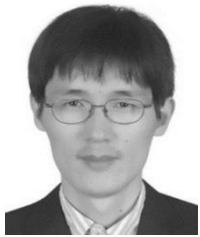

**Yan Hu** received the BSc and PhD degrees in computer science from University of Science and Technology of China (USTC), China, in 2002 and 2007, respectively. He is currently an assistant professor at School of Software, Dalian University of Technology, China. His research interests include model checking, program analysis, and software engineering. He is a member of the ACM and the CCF.

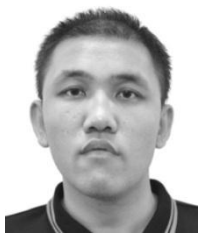

**Zhilei Ren** received the BSc degree in software engineering in 2007 and the PhD degree in 2013, from Dalian University of Technology, China. He is currently a postdoctoral researcher of School of Software at Dalian University of Technology. His research interests include evolutionary computation and its applications in software engineering. He is a member of the ACM and the CCF.

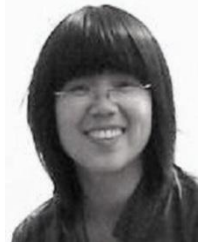

**Weiqin Zou** received the BSc degree in software engineering in 2010 and the MSc degree in computer application and technology in 2013, from Dalian University of Technology. She is currently a teaching assistant of Department of Information Engineering, Jiangxi University of Science and Technology, China. Her research interests include mining software repositories and machine learning.

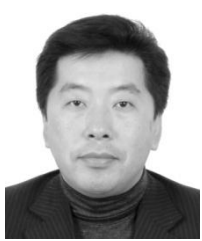

**Zhongxuan Luo** is a professor of School of Mathematical Sciences at Dalian University of Technology, China. He received the BSc, MSc, and PhD degrees in computational mathematics from Jilin University (China) in 1985, Jilin University in 1988, and Dalian University of Technology in 1991, respectively. His research interests include multivariate approximation theory and computational geometry.

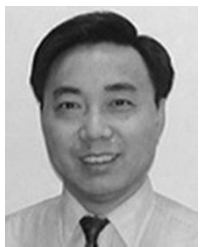

**Xindong Wu** is a Yangtze River Scholar in the School of Computer Science and Information Engineering at the Hefei University of Technology (China), a professor of the Department of Computer Science at the University of Vermont (USA), and a Fellow of the IEEE and AAAS. He holds a PhD degree in Artificial Intelligence from the University of Edinburgh, Britain. His research interests include data mining, knowledge-based systems, and Web information exploration.